\documentclass[prd, aps, superscriptaddress, preprintnumbers, twocolumn, floatfix, nofootinbib]{revtex4}
\pdfoutput=1

\usepackage{amsfonts}
\usepackage{amsmath}
\usepackage{amssymb}
\usepackage{bm}
\usepackage{dcolumn}
\usepackage{graphicx}   
\usepackage{graphics}
\usepackage[latin1]{inputenc}
\usepackage{latexsym}
\usepackage{rotating}
\usepackage[colorlinks]{hyperref}
\hypersetup{linkcolor=blue,citecolor=blue,urlcolor=red}
\usepackage{graphicx}
\usepackage[dvipsnames]{xcolor}
\usepackage{soul}

\newcommand\be{\begin{equation}}
\newcommand\ba{\begin{eqnarray}}
\newcommand\ee{\end{equation}}
\newcommand\ea{\end{eqnarray}}

\begin{document}

\title{Solution of the Size and Horizon Problems from Classical String Geometry}

\author{Heliudson Bernardo}
\email{heliudson.deoliveirabernardo@mcgill.ca}
\affiliation{Department of Physics, McGill University, Montr\'{e}al, QC, H3A 2T8, Canada}

\author{Robert Brandenberger}
\email{rhb@physics.mcgill.ca}
\affiliation{Department of Physics, McGill University, Montr\'{e}al, QC, H3A 2T8, Canada}

\author{Guilherme Franzmann}
\email{guilherme.franzmann@su.se}
\affiliation{Nordita, KTH Royal Institute of Technology and Stockholm University, 
Roslagstullsbacken 23, SE-106 91 Stockholm, Sweden}

\date{\today}

\begin{abstract}

 In a recent paper we developed a string cosmology background from
       classical string geometry. Here, we show that this background yields a
       solution to the size and horizon problems of Standard Big Bang cosmology while remaining compatible with the Transplanckian Censorship Conjecture.
       We also take a first look at the evolution of cosmological perturbations in
       this model. 

\end{abstract}

\maketitle


\section{Introduction}
\label{sec:intro}

If superstring theory is the correct framework which unifies all forces of nature at the quantum level, then strings must have played a crucial role at the high densities of the early universe. A theory based on one-dimensional strings as the fundamental objects has new symmetries and new degrees of freedom which theories based on point particles do not have. In particular, strings do not only have momentum modes, but also oscillatory and winding modes. Associated with the existence of winding modes is a new symmetry, namely the T-duality symmetry \cite{Green:1982sw,Kikkawa:1984cp,Sakai:1985cs,Alvarez:1994dn,Boehm:2002bm} which implies that, for a compact background (e.g. a toroidal background), physics on a space of radius $R$ is equivalent to physics on a space of radius $1/R$ (in string units). From the partition function of a gas of closed strings \cite{Deo:1989bv} it then follows that the temperature $T(R)$ obeys this symmetry which implies that the temperature of a gas of closed strings is finite \cite{Brandenberger:1988aj} for any radius and always below the Hagedorn temperature \cite{Hagedorn:1965st}. 

String Gas Cosmology (SGC) \cite{Brandenberger:1988aj} (see also \cite{Kripfganz:1987rh}) is a scenario based on the above results in which an emergent phase dominated by a gas of strings with temperature close to $T_H$ precedes the Standard Big Bang expansion. To attain thermal equilibrium, a quasi-static phase in Einstein frame is postulated, with a nearly constant scale factor. Regarding cosmological perturbations, it was shown in \cite{Nayeri:2005ck} that thermal fluctuations yield a nearly scale-invariant spectrum of scalar and tensor perturbations, with a red tilt for the former and blue tilt for the latter \cite{Brandenberger:2014faa, Brandenberger:2006xi}, and non-Gaussianities which are Poisson-suppressed on large scales \cite{Chen:2007js}. Hence, String Gas Cosmology explains the origin of structure of the universe and it is an alternative to the inflationary paradigm (for reviews on SGC, see \cite{Brandenberger:2011et,Brandenberger:2008nx,Battefeld:2005av}).

At late times, once Einstein gravity may be assumed as a low energy field theory, the interplay between momentum and winding modes naturally stabilizes the size moduli of extra dimensions  \cite{Patil:2004zp,Patil:2005fi,Watson:2003gf,Watson:2004aq} while the shape moduli are also stabilized by string flux effects \cite{Brandenberger:2005bd}. In \cite{Danos:2008pv} it was shown that non-perturbative gaugino condensation can be used to stabilize the dilaton, with no interference on the other moduli stabilization, and the same non-perturbative effect was shown to break supersymmetry at a high scale \cite{Mishra:2011fc}.

Apart from the successes of SGC, it had remained an open problem until recently to justify the assumption of an early quasi-static phase for the scale factor in the Einstein frame in the high temperature regime. In fact, it was known that the Einstein equations could not be used since they are not consistent with the T-duality symmetry of string theory. One could have hoped that {\it Pre-Big-Bang Cosmology} \cite{Gasperini:1992em} (see also \cite{Tseytlin:1991xk}, and \cite{Gasperini:2002bn} for a review) could tackle the issue since it is compatible with T-duality, but without higher order string corrections ($\alpha'$ corrections) the individual solutions are neither quasi-static nor non-singular. 

{\it Double Field Theory} (DFT) \cite{Duff:1989tf,Tseytlin:1990hn,Kugo:1992md,Siegel:1993th, Hull:2009mi} (see e.g. \cite{Aldazabal:2013sca} for a review) is a proposal for an effective field theory description of space-time which is built on the T-duality symmetry of string theory. DFT is a field theory which describes a {\it doubled space}, with an extra set of spatial coordinates dual to the regular ones. The original set can be viewed \cite{Brandenberger:1988aj} as the Fourier transform of the string momentum modes, and the dual set as the Fourier transform of the string winding modes. Thus, the T-duality symmetry of string theory is manifest. Point particle motion in DFT was studied in \cite{Brandenberger:2017umf}, where it was shown that geodesics can be extended infinitely far into both the past and the future. In \cite{Brandenberger:2018xwl,Brandenberger:2018bdc,Bernardo:2019pnq}, cosmological solutions of the DFT equations of motion were studied in the presence of a matter source corresponding to a gas of strings, and it was shown that the solutions are non-singular. However, the equations of motion contained neither string nor quantum corrections, and the solutions obtained did not have a quasi-static phase at high energy densities. We expect that at high densities higher order corrections in $\alpha^{\prime}$ become crucial. 

In an important paper \cite{Hohm:2019jgu}, Hohm and Zwiebach recently were able to classify all higher order $\alpha^{\prime}$ correction terms to the effective field theory action consistent with the $O(d,d)$ symmetry of homogeneous and isotropic backgrounds. In a followup paper \cite{Bernardo:2019bkz}, the analysis was extended to include matter sources. The formalism established in \cite{Hohm:2019jgu,Bernardo:2019bkz} thus provides the framework to study classical superstring cosmology \footnote{Classical since the quantum corrections are not taken into account.} at the mini-superspace level including all possible $\alpha^{\prime}$ corrections, in the presence of matter. Some solutions and their stability were discussed in \cite{Bernardo:2020zlc}.

In a recent paper \cite{Bernardo:2020nol} we studied solutions of the classical string equations of \cite{Bernardo:2019bkz} which begin as a de Sitter phase in the String frame. If matter is dominated by winding modes, the Einstein frame metric is static. Thus, these solutions can be viewed as candidates for the {\it emergent universe} \cite{Brandenberger:2019jbs} which SGC posits. In this paper we study some phenomenological consequences of these solutions. In particular, we show that the size and horizon problems of Standard Big Bang cosmology can be solved. The scenario can be made consistent with the {\it Trans-Planckian Censorship Conjecture} (TCC) \cite{Bedroya:2019snp}. We also take a first look at the evolution of cosmological perturbations in this framework.

In the following we will work in a $d + 1$ dimensional space-time with $n$ internal and $d - n$ external dimensions (we will consider $d = 9$ and $n = 6$). We use units in which $c = \hbar = 1$, and denote $\kappa^2 = 8 \pi G$, with $G$ being Newton's gravitational constant. The equation of state parameter $w$ is the ratio of pressure to energy density. Time is denoted by the variable $t$, temperature by $T$, and scale factor by $a$. Subscripts $0$ on these quantities denote the values of these variables at the present time.

\section{Review of $\alpha'$-Cosmology and summary of the model}

The basic variables for classical string cosmology can be expressed in terms of the matrix
\be
S \, = \, {\begin{bmatrix} b g^{-1} & g - b g^{-1} b \\
g^{-1} & - g^{-1} b 
\end{bmatrix}}
\, ,
\ee
where $g$ is the spatial metric, a symmetric $d \times d$ matrix, and $b$ is the antisymmetric tensor field. Together with the dilaton $\phi$, these are the massless modes of bosonic string theory.

In \cite{Hohm:2019jgu, Bernardo:2019bkz} it was shown that at the minisuperspace level, when the spatial metric can be written as
\be
g_{ij} \, = \, a(t)^2 \delta{ij} \, ,
\ee
where $a(t)$ is the cosmological scale factor,  the most general action consistent with $O(d,d)$ symmetry can be written in the form
\ba
S \, &=& \, \frac{1}{2 \kappa^2} \int d^dx dt e^{-\Phi} 
\bigl[ - ({\cal{D}}\Phi)^2 + X({\cal{D}}S) \bigr] \nonumber \\
&+& \, S_m(\Phi, n, S, \chi) \, ,
\ea
where
\be
\Phi \, = \, 2 \phi - {\rm{ln}}(\sqrt{g}) 
\ee
is the shifted dilaton, a scalar under $O(d,d)$ transformations, ${\cal{D}}$ indicates the covariant derivative operator, $X(f)$ is a scalar function which does not involve higher derivatives of $f$, $n$ is the lapse function, and $\chi$ stand for the matter fields. The first line in the above is the space-time action, the second line is the matter action. The function $X(f)$ can be expanded in powers of $f$, with coefficients which are not constrained by the $O(d,d)$ symmetry (but which could be determined from full string theory).

In \cite{Bernardo:2020nol}, solutions of the equations of motion were identified in which all nine spatial dimensions start out with a size smaller than the string length. In this case, the equation of state of the string gas is expected to be dominated by the winding modes, and the solution correspond to exponential expansion in the String frame but yield a static metric in the Einstein frame. This initial phase was denoted {\it Stage 1} in \cite{Bernardo:2020nol}. 

\begin{figure}[h!]
    \centering
    \includegraphics[scale=.39]{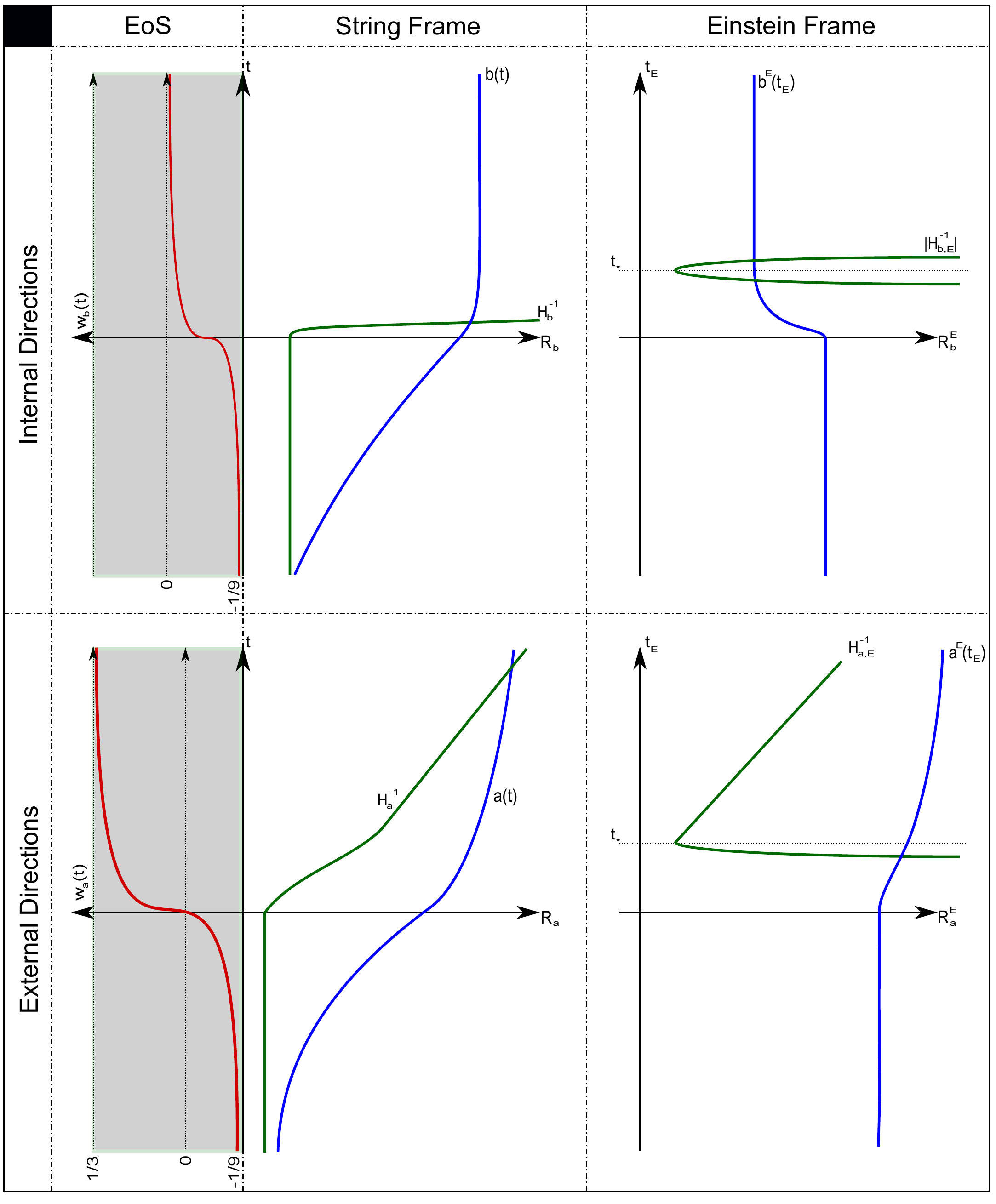}
    \caption{Sketch of the equations of state and Hubble radii in the string and Einstein frames as a function of time (vertical axis), for both internal and external directions. Stage 1 lasts until the origin of the time axis. The time $t_{*}$ corresponds to the end of Stage 3 when the dilaton stabilizes. Most of the time interval between $t = 0$ and $t = t_{*}$ corresponds to Stage 3.}
    \label{fig:model}
\end{figure}

When the size of space approaches the string scale, winding modes start to annihilate into momentum and oscillatory string states. However, as discussed in \cite{Brandenberger:1988aj}, this annihilation only happens in three of the spatial dimensions, while the other six remain wrapped by winding modes. The equation of state evolves towards $w = 1/3$ for the three large dimensions (``external'') and $w = 0$ for the six ``internal'' dimensions. This is a short phase, denoted \textit{Stage 2}, during which the internal directions contract while the external ones undergo super-exponential expansion (in the Einstein frame). As the equation of state parameter in the external directions approaches the radiation form $w = 1/3$, the internal dimensions become static, while the external ones continue with super-exponential expansion. This is \textit{Stage 3} which we study here. This phase ends once the dilaton becomes stabilized by non-perturbative effects. The evolution of the equations of state and the radii in both String and Einstein frames is depicted in Figure \ref{fig:model} for both internal and external directions.

\section{Solving the Size and Horizon Problems of Standard Cosmology}

In this section we show that the horizon and size problems of Standard Big Bang cosmology can be solved with the classical string background which we have described. Both problems can be solved if we can show that the comoving scale corresponding to the current Hubble radius originates at the beginning of Stage 3 with a physical wavelength smaller than the Hubble radius. In this case, it follows that one initial Hubble patch can grow into the observed universe, and as a corollary it follows that the region over which we observe the Cosmic Microwave Background (CMB) to be isotropic also starts out smaller than an initial Hubble patch, thus allowing for a causal mechanism to produce the observed isotropy. 

During Stage 3, the evolution of the dilaton and of the Einstein frame Hubble parameter are given by \cite{Bernardo:2020nol}
\begin{widetext}
\begin{align}
    \phi^{(3)}(t_E) &= \phi^{(3)}(t_r) - \frac{d-1}{2}\ln\left[1 - e^{\frac{2\phi^{(3)}(t_r)}{d-1}}\frac{(d-n+1)H^{(3)}}{d-1}(t_E - t_{E,0})\right], \nonumber\\
    H_{E}^{(3)}(t_E) &= \frac{e^{\frac{2\phi^{(3)} (t_r)}{d-1}}}{d-1}H^{(3)}(n-2)\left[1 - e^{\frac{2\phi^{(3)}(t_r)}{d-1}}\frac{(d-n+1)}{d-1}H^{(3)}(t_E - t_E(t_r))\right]^{-1}, \label{super_expon}
\end{align}
\end{widetext}
where $t_{E,0} \equiv t_E(t_r)$ and $t_r$ denotes the beginning of stage 3 in the String frame.

Notice that there is a bound on the maximum duration of Stage 3 in the Einstein frame, which corresponds to the time the Hubble parameter becomes singular,
\begin{equation}\label{Stage3maxdurationinEF}
   \Delta t_{E, \text{max}} \equiv t_{E, \text{max}} - t_{E, 0} =\frac{1}{H_I}\frac{(n-2)}{(d+1-n)} = \frac{1}{H_I},
\end{equation}
where the last equality holds after considering $d=9$, $n=6$, and we have defined
\begin{equation}
    H_I \equiv \frac{e^{\frac{2\phi^{(3)} (t_r)}{d-1}}}{d-1}H^{(3)}(n-2) = \frac{e^{\frac{\phi^{(3)} (t_r)}{4}}}{2}H^{(3)}.
\end{equation}
Using this definition of $H_I$, and defining $t_I \equiv t_{E, 0}$ we can rewrite the evolution of the dilaton and the Hubble parameter for $d = 9$ and $n = 6$ in the simpler form
\ba
    \phi^{(3)}(t_E) \, &=& \, \phi^{(3)}(t_r) - 4 \ln\left[1 - H(t_E - t_I)\right], \nonumber\\
    H_{E}^{(3)}(t_E) \, &=& \, H_I \left[1 - H_I(t_E - t_I)\right]^{-1}. \label{super_expo}
\ea
Recall that the maximum duration of this stage in the Einstein frame corresponds to the String frame time variable $t$ going from $0$ to $\infty$. Thus, although from the Einstein frame point of view the finite $t_{E,\text{max}}$ corresponds to the time where we get to a singularity, in the String frame this happens only at infinity. Nonetheless, in order to avoid the singularity we impose this phase of Stage 3 to end before reaching $\Delta t_{E,\text{max}}$.

Before we proceed to investigate a solution to the horizon problem, let us remember that the Stage 3 was defined to end when the dilaton enters in the strong string coupling regime, $\phi^{(3)}(t_*) \sim 0$, so that $g_s < 1$ throughout Stage 3. Hence, from the dilaton's evolution in the String frame (eq. (58) in \cite{Bernardo:2020nol}) we have
\begin{equation}
    0 \, = \, \phi^{(3)}(t_*)= \phi^{(3)}(t_r) + 2 H^{(3)}(t_*-t_r),
\end{equation}
which implies a bound on the duration of Stage 3 in the String frame
\begin{equation}\label{Stage3durationinSF}
    \Delta t \, < \, -\frac{\phi^{(3)}(t_r)}{2H^{(3)}} \, ,
\end{equation}
which depends on the string coupling at the beginning of this stage, $\phi^{(3)}(t_r)$.

Let us translate the bound (\ref{Stage3durationinSF}) to the Einstein frame. From equation (59) in \cite{Bernardo:2020nol}, we have
\be \label{strongcouplingconstraint}
    \Delta t_E \, < \, \frac{1}{H_I}\left(1- e^{\frac{\phi^{(3)}(t_r)}{4}}\right),
\ee
and we see that this is compatible with (\ref{Stage3maxdurationinEF}). Summarizing, there is an upper bound on the duration of Stage 3 in the String frame which is compatible with the maximum possible duration of this phase in the Einstein frame. 

Let us now investigate if there could be enough accelerated expansion to solve the horizon problem (this will then also solve the size problem). We demand that our current observable Universe should fit in the comoving Hubble radius at the beginning of the phase of accelerated expansion, i.e. (neglecting the evolution during Stage 2)
\begin{equation}\label{CC1}
H_0^{-1} \ \leq \, \frac{a_0}{a_I} H_I^{-1} \, .
\end{equation}
In terms of the end time $t_R$ of Stage 3 in the Einstein frame, when the scale factor is $a_R$ and then Einstein frame Hubble parameter takes the value $H_R$, this can be re-written as
\be \label{CC2}
H_0^{-1} \, \leq \, \frac{a_0}{a_R} \frac{a_R}{a_I} \frac{H_R}{H_I} H_R^{-1}. \,
\ee

From (\ref{super_expo}) it follows that the Einstein frame scale factor becomes 
\be \label{SFevol}
a_E(t_E) \, = \, a_I[1 - H_I(t_E-t_I)]^{-1} \, ,
\ee
where $a_I$ is the value of the scale factor at the beginning of Stage 3. Hence, during Stage 3 we take $H \sim a$, and thus (\ref{CC2}) becomes the following condition on the duration of this stage:
\ba \label{CC3}
\left( \frac{a_R}{a_I} \right)^2 \, &\geq& \, \frac{H_R}{H_0} \frac{T_0}{T_R} \\
&=& \, z_{eq}^{-1/2} \frac{T_R}{T_0} \nonumber \, .
\ea
Here, we have made the approximation that the universe becomes dominated by radiation immediately after $t_R$ (and we neglected the short intermediate phase during which the scaling $H \sim t^{-1}$ is achieved). Similarly, we have neglected the effects of dark energy at the present time, and we have hence taken the universe to be dominated by radiation from $t_R$ until $t_{eq}$ (with associated redshift $z_{eq}$), and then dominated by pressureless matter. Thus, we can use the scaling $T \sim a^{-1}$ for $t > t_R$, which corresponds to the constancy of the entropy density, and $H \sim t^{-1}$. Furthermore, $T \sim t^{-2/3}$ for $t > t_{eq}$  and $T \sim t^{-1/2}$ for $t_R < t < t_{eq}$, and this is used to obtain the second line. The scaling $T \sim a^{-1}$ applies also during Stage 3, and this allows us to replace $T_R$ in (\ref{CC3}) by the initial temperature $T_I$ at the beginning of Stage 3, a temperature which we expect to be given by the string scale. Thus, (\ref{CC3}) becomes
\be \label{CC4}
\left( \frac{a_R}{a_I} \right)^3 \, \geq \, \frac{T_I}{T_0} z_{eq}^{-1/2} \, \equiv \, \gamma \, .
\ee
Inserting for $T_I$ the string scale taken to be $T_I \sim 10^{16}$ GeV, we find that the right hand side of (\ref{CC4}) is of the order $\gamma \sim 10^{26}$. Thus, provided that (\ref{CC4}) is satisfied, our model can solve the horizon and size problems of Standard Big Bang cosmology.

The condition (\ref{CC4}) on the duration of Stage 3 sets an upper bound on the initial value of the dilaton field at the beginning of Stage 3. Inserting the formula (\ref{SFevol}) for the evolution of the scale factor during Stage 3 we see that (\ref{CC4}) takes the form
\be
1 - H_I(t_R - t_I) \, < \, \gamma^{-1/3} \, ,
\ee
and, using the evolution of the dilaton from (\ref{super_expo}), this becomes a lower bound on the change $\Delta \phi$ of the dilaton between times $t_I$ and $t_R$:
\be
e^{- \Delta \phi / 4} \, \leq \, \gamma^{-1/3} \, ,
\ee
which yields an upper bound on the initial value $\phi_I$ of the dilaton. Assuming that $\phi(t_R) = 0$ we find
\be \label{final}
\phi_I \, < - \frac{4}{3} \ln \gamma \,,
\ee
which is compatible with the weak coupling assumption for neglecting $g_s$ corrections.

\section{Connection with the Trans-Planckian Censorship Conjecture}

Recently \cite{Bedroya:2019snp}, a criterion on viable effective field theory cosmologies has been conjectured. It states that no scale which had an initial wavelength smaller than the Planck length could ever have exited the Hubble radius. This {\it Trans-Planckian Censorship} conjecture (TCC) shields a late time observer from being affected by modes which were initially trans-Planckian. The reason why the Hubble horizon is the relevant scale is because fluctuations only oscillate on sub-Hubble scales, while they freeze out and increase in amplitude once they exit the Hubble horizon. If the fluctuations begin as quantum vacuum perturbations, they become squeezed vacuum states on super-Hubble scales, they can decohere due to intrinsic nonlinearities in the system, and they classicalize \cite{Kiefer:1998qe}. The TCC shields the observer from classical perturbations which emerge in the trans-Planckian ``sea'', a region where effective field theory cannot be trusted.

The TCC implies stringent constraints \cite{Bedroya:2019tba} on standard inflationary models, only allowing low scale inflation ($V^{1/4} < 10^{10} {\rm{GeV}}$) with an utterly negligible amplitude of primordial gravitational waves. Since our model is also an effective field theory which does not include all string degrees of freedom, it is interesting to check whether Stage 3 satisfies the TCC or not.

The TCC bound implies that the Planck length at the initial time $t_I$ can never become larger than the Hubble radius at any later time $t_F$:
\begin{equation}
    \frac{a_F}{a_I} \, < \, \frac{M_{\text{Pl}}}{H_F} \, .
\end{equation}
When applied to Stage 3, the TCC implies a bound to its duration $\delta t$
\begin{equation}
    \delta t \, < \, \frac{1}{H_I}\left(1- \frac{H_F}{M_{\text{Pl}}}\right),
\end{equation}
which is consistent with the upper bound (\ref{Stage3maxdurationinEF}). Imposing the stronger bound (\ref{strongcouplingconstraint}) we have
\begin{equation}
    e^{\phi^{(3)}(t_r)/4} \, > \, \frac{H_F}{M_{\text{Pl}}}.
\end{equation}
Thus, if we want the Stage 3 to solve the horizon problem \emph{and} to satisfy the TCC, we need
\begin{equation}
    \gamma^{-4/3} \, > \,  e^{\phi^{(3)}(t_r)} > \left(\frac{H_F}{M_{\text{Pl}}}\right)^4,
\end{equation}
which yields a bound to $H_F$,
\begin{equation}
    H_F \, < \, M_{\text{Pl}}\gamma^{-1/3}.
\end{equation}
This is a mild constraint compared to the constraint which the TCC imposes \cite{Bedroya:2019tba} on slow-roll inflationary models. In particular for $\gamma \sim 10^{26}$ the upper bound implies that the energy scale at which Stage 3 ends should not be greater than $\sim 10^{10} \text{GeV}$.

\section{A First Look at the Evolution of Cosmological Perturbations}


In this section we will take a first look at the evolution of linear cosmological fluctuations in the background discussed in previous sections. To study cosmological fluctuations it is useful to work in the Einstein frame. At linear level, all Fourier modes evolve independently. We are interested in fluctuations of the metric in the external space which are on cosmological scales today. Since Stage 1 is static in the Einstein frame (and since Stage 2 is of very short duration), the modes must have exited the Hubble horizon early in Stage 3.  In this preliminary study we will follow the evolution of the fluctuations making use of the perturbative Einstein equations, to which the full classical string equations will reduce in the infrared limit. A more consistent study would require us to understand all $\alpha^{\prime}$ corrections beyond the minisuperspace approximation, something we do not have a handle on yet.

We will follow the fluctuations of the Einstein frame metric. In longitudinal gauge, the external metric including linearized cosmological fluctuations can be written in the form (see \cite{Mukhanov:1990me} for a comprehensive review article)
\be
ds^2 \, = \, a(\tau)^2 \bigl( - (1 + 2\Phi) d\tau^2 + (1 - 2\Phi) d{\bf{x}}^2 \bigr) \, ,
\ee
where $\Phi({\bf{x}}, \tau)$ is the gravitational potential, and $\tau$ is conformal time. We are not considering vector and tensor fluctuations. At linear order they decouple from the scalar cosmological perturbations which we are considering here.

The total action can be expanded to quadratic order in terms of the amplitude of the fluctuations. Since matter fluctuations induce scalar metric fluctuations via the generalized Poisson equations, the scalar sector of the fluctuations can be described in terms of a single canonically normalized field $v$ \cite{Sasaki:1986hm,Mukhanov:1988jd} which - for perfect fluid matter - is the following combination of metric and matter fluctuations:
\be
v \, = \, \frac{1}{\sqrt{2} l_{pl}} \left( \varphi_v + \frac{2 a \beta^{1/2}}{{\cal{H}}c_s} \Phi \right)
\, ,
\ee
where $\varphi_v$ is the velocity potential of the fluid, $c_s$ is its speed of sound (which we set to $1$ in the following), $l_{pl}$ is the Planck length, ${\cal{H}}$ is the Hubble expansion rate in conformal time, and
\be
\beta \, = \, {\cal{H}}^2 - {\cal{H}}^{\prime} \, ,
\ee
where the prime denotes the derivative with respect to conformal time.

In Fourier space, the resulting equation of motion is
\be
v_k^{\prime \prime} + \left(k^2 - \frac{a^{\prime \prime}}{a} \right)v_k \, = \, 0 \, ,
\ee
where $k$ is the comoving momentum, and a prime indicates the derivative with respect to conformal times $\tau$. In the fully $\alpha'$-corrected theory, we expect modifications in the squeezing mass term of this equation and in the canonical variable that makes the friction term vanish. In an expanding background, it follows from the above that $v_k$ oscillates on sub-Hubble scales, i.e. for
\be
k \, > \, {\cal{H}} \, \equiv \, \frac{a^{\prime}}{a} \, .
\ee
In an expanding background like the one we are considering, the dominant mode on super-Hubble scales is growing as
\be \label{growmode}
v_k \, \propto \, a \, .
\ee

We are interested in studying the power spectrum
\be
P_v(k, t) \, \equiv \, k^3 |v_k|^2(t) \, 
\ee
of the fluctuation variable $v$. For a scale-invariant spectrum, $P_v(k, t)$ is independent of $k$. More generally, we can introduce the scalar spectral index $n_s$ via
\be
P_v(k, t) \, \sim \, k^{n_s - 1} \, ,
\ee
where scale-invariance corresponds to $n_s = 1$. 

\begin{figure}[h!]
    \centering
    \includegraphics[scale=.6]{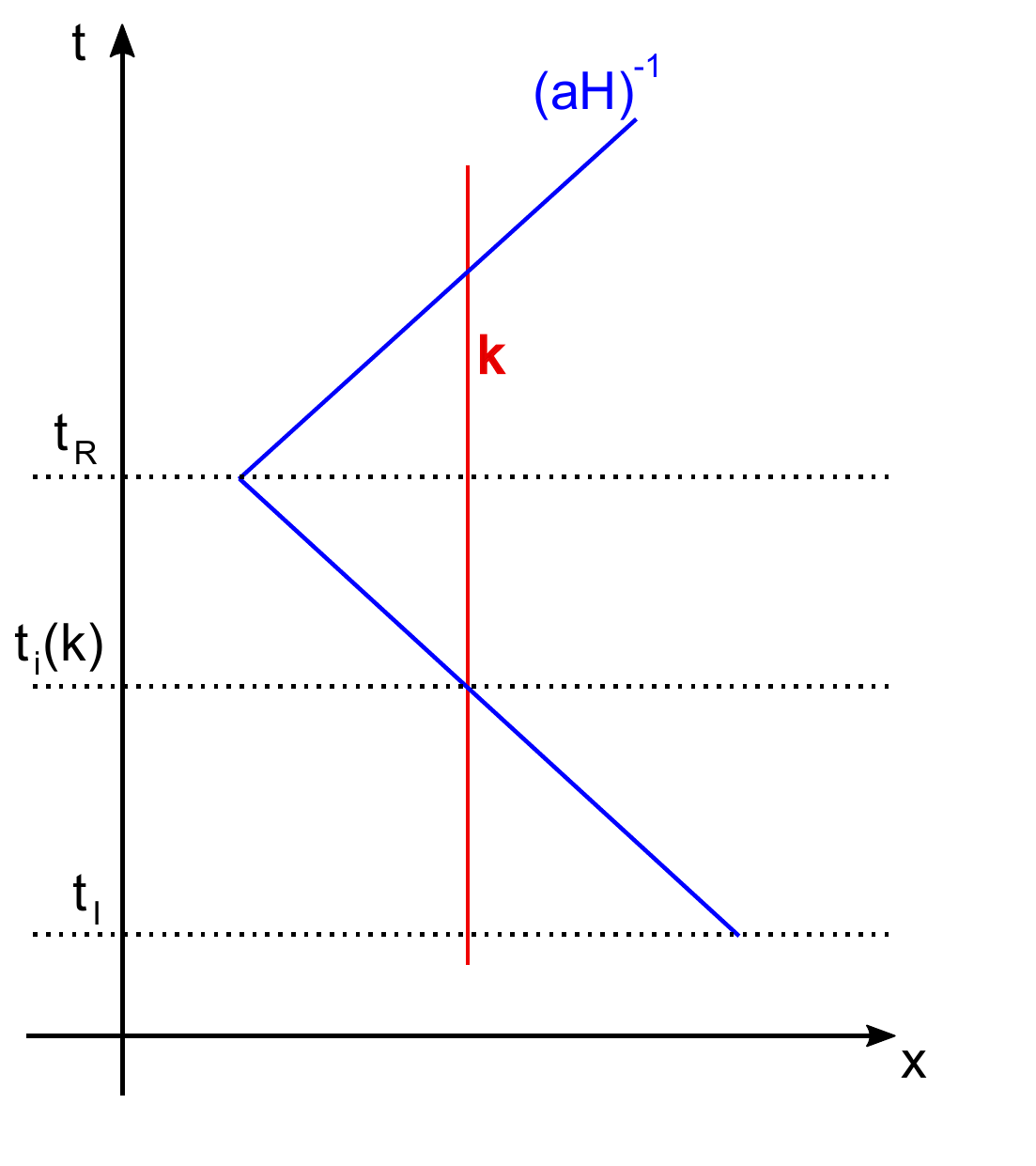}
    \caption{Space-time sketch (with the horizontal axis depicting comoving spatial coordinate and the vertical axis being time) showing the evolution of the comoving Hubble radius $(aH)^{-1}$  during and after Stage 3 (beginning at time $t_I$ and ending at time $t_R$) of our model. The wavelength of a fluctuation mode (the vertical line labelled by $k$) exits the Hubble radius at a time $t_i(k)$ and then re-enters at some late time.}
    \label{BBFfig}
\end{figure}

Consider a fluctuations mode which exits the Hubble radius at time $t_i(k)$ during Stage 3 (see Fig. \ref{BBFfig}). We wish to compute the spectrum of $v$ at the end of Stage 3 ($t = t_R$) as a function of the spectrum at the beginning of Stage 3 ($t = t_I$). Since the fluctuations oscillate on sub-Hubble scales and evolve as (\ref{growmode}) on super-Hubble scales, we have
\be
P_v(k, t_R) \, = \, \left( \frac{a(t_R)}{a(t_i(k))} \right)^2 P_v(k, t_I) \, .
\ee
The Hubble radius crossing time $t_i(k)$ is given by
\be
a(t_i(k)) k^{-1} \, = \, H_E(t_i(k))^{-1} \, .
\ee
Making use of the equations for $a(t)$ and $H_E(t)$ we obtain
\be
a(t_i(k)) \, = \, a(t_I) \left[ \frac{k}{a(t_I) H_I} \right]^{1/2} \, .
\ee
Hence, the initial power spectrum at the time $t_I$ gets boosted in the infrared by a factor proportional to $k^{-1}$ (this is different as compared to slow-roll inflation, where the boost factor is proportional to $k^{-2}$):
\be
P_v(k, t_R) \, = \, \left( \frac{a(t_R)}{a(t_I)} \right)^2 \frac{H_I a(t_I)}{k} P_v(k, t_I) \, .
\ee
In order to obtain a scale-invariant spectrum of curvature fluctuations at late times, an initial spectrum at the end of Stage 2 with power spectrum proportional to $k$ is required. In work in progress we are investigating the origin of fluctuations in our model.

\section{Summary and Conclusions}

We have explored some aspects of the cosmology of the classical string background discovered in \cite{Bernardo:2020nol}. This background is a candidate for an {\it emerging cosmology}. The evolution begins with all nine dimensions of space smaller than the string scale, and matter being a gas of strings dominated by winding modes. In this case, the equations of motion for a classical string background admit a solution in which all dimensions of space begin by expanding exponentially in the String frame, while they are static in the Einstein frame (Stage 1). When the size of the spatial dimensions approaches the string scale, the equation of state of the string gas begins to deviate from that of pure winding number domination. Since winding modes can only efficiently annihilate in three spatial dimensions, our background evolves into one in which three dimensions grow while the others remain compact. After a brief intermediate phase (Stage 2), the internal dimensions become static while the external ones undergo super-exponential expansion (Stage 3). Stage 3 ends once the dilaton is stabilized.

We have shown that as a consequence of the super-exponential expansion of the external dimensions, our background can solve the horizon and size problems of Standard Big Bang cosmology, provided that the initial value of the dilaton is sufficiently small such that Stage 3 lasts sufficiently long. We have shown that our background can satisfy the {\it Trans-Planckian Censorship Conjecture}. We have also studied the evolution of cosmological fluctuations during Stage 3, and found that the power spectrum of fluctuations is bootsed in the infrared by a factor proportional to $k^{-1}$, where $k$ is the comoving momentum. 

\section*{Acknowledgments}

The research at McGill is supported, in part, by funds from NSERC and from the Canada Research Chair program. RB thanks the  Pauli Center and the Institutes for Theoretical Physics and of Particle Astrophysics of the ETH for hospitality.

\bibliographystyle{bibstyle} 
\bibliography{references}

\providecommand{\href}[2]{#2}\begingroup\raggedright\begin{thebibliography}{10}

\bibitem{Green:1982sw}
M.~B. Green, J.~H. Schwarz and L.~Brink, \emph{{N=4 Yang-Mills and N=8
  Supergravity as Limits of String Theories}},
  \href{https://doi.org/10.1016/0550-3213(82)90336-4}{\emph{Nucl. Phys. B}
  {\bfseries 198} (1982) 474--492}.

\bibitem{Kikkawa:1984cp}
K.~Kikkawa and M.~Yamasaki, \emph{{Casimir Effects in Superstring Theories}},
  \href{https://doi.org/10.1016/0370-2693(84)90423-4}{\emph{Phys. Lett. B}
  {\bfseries 149} (1984) 357--360}.

\bibitem{Sakai:1985cs}
N.~Sakai and I.~Senda, \emph{{Vacuum Energies of String Compactified on
  Torus}}, \href{https://doi.org/10.1143/PTP.75.692}{\emph{Prog. Theor. Phys.}
  {\bfseries 75} (1986) 692}.

\bibitem{Alvarez:1994dn}
E.~Alvarez, L.~Alvarez-Gaume and Y.~Lozano, \emph{{An Introduction to T duality
  in string theory}},
  \href{https://doi.org/10.1016/0920-5632(95)00429-D}{\emph{Nucl. Phys. B Proc.
  Suppl.} {\bfseries 41} (1995) 1--20},
  [\href{https://arxiv.org/abs/hep-th/9410237}{{\ttfamily hep-th/9410237}}].

\bibitem{Boehm:2002bm}
T.~Boehm and R.~Brandenberger, \emph{{On T duality in brane gas cosmology}},
  \href{https://doi.org/10.1088/1475-7516/2003/06/008}{\emph{JCAP} {\bfseries
  06} (2003) 008}, [\href{https://arxiv.org/abs/hep-th/0208188}{{\ttfamily
  hep-th/0208188}}].

\bibitem{Deo:1989bv}
N.~Deo, S.~Jain and C.-I. Tan, \emph{{STRING STATISTICAL MECHANICS ABOVE
  HAGEDORN ENERGY DENSITY}},
  \href{https://doi.org/10.1103/PhysRevD.40.2626}{\emph{Phys. Rev. D}
  {\bfseries 40} (1989) 2626}.

\bibitem{Brandenberger:1988aj}
R.~H. Brandenberger and C.~Vafa, \emph{{Superstrings in the Early Universe}},
  \href{https://doi.org/10.1016/0550-3213(89)90037-0}{\emph{Nucl. Phys. B}
  {\bfseries 316} (1989) 391--410}.

\bibitem{Hagedorn:1965st}
R.~Hagedorn, \emph{{Statistical thermodynamics of strong interactions at
  high-energies}}, {\emph{Nuovo Cim. Suppl.} {\bfseries 3} (1965) 147--186}.

\bibitem{Kripfganz:1987rh}
J.~Kripfganz and H.~Perlt, \emph{{Cosmological Impact of Winding Strings}},
  \href{https://doi.org/10.1088/0264-9381/5/3/006}{\emph{Class. Quant. Grav.}
  {\bfseries 5} (1988) 453}.

\bibitem{Nayeri:2005ck}
A.~Nayeri, R.~H. Brandenberger and C.~Vafa, \emph{{Producing a scale-invariant
  spectrum of perturbations in a Hagedorn phase of string cosmology}},
  \href{https://doi.org/10.1103/PhysRevLett.97.021302}{\emph{Phys. Rev. Lett.}
  {\bfseries 97} (2006) 021302},
  [\href{https://arxiv.org/abs/hep-th/0511140}{{\ttfamily hep-th/0511140}}].

\bibitem{Brandenberger:2014faa}
R.~H. Brandenberger, A.~Nayeri and S.~P. Patil, \emph{{Closed String
  Thermodynamics and a Blue Tensor Spectrum}},
  \href{https://doi.org/10.1103/PhysRevD.90.067301}{\emph{Phys. Rev. D}
  {\bfseries 90} (2014) 067301},
  [\href{https://arxiv.org/abs/1403.4927}{{\ttfamily 1403.4927}}].

\bibitem{Brandenberger:2006xi}
R.~H. Brandenberger, A.~Nayeri, S.~P. Patil and C.~Vafa, \emph{{Tensor Modes
  from a Primordial Hagedorn Phase of String Cosmology}},
  \href{https://doi.org/10.1103/PhysRevLett.98.231302}{\emph{Phys. Rev. Lett.}
  {\bfseries 98} (2007) 231302},
  [\href{https://arxiv.org/abs/hep-th/0604126}{{\ttfamily hep-th/0604126}}].

\bibitem{Chen:2007js}
B.~Chen, Y.~Wang, W.~Xue and R.~Brandenberger, \emph{{String Gas Cosmology and
  Non-Gaussianities}}, {\emph{The Universe} {\bfseries 3} (2015) 2--10},
  [\href{https://arxiv.org/abs/0712.2477}{{\ttfamily 0712.2477}}].

\bibitem{Brandenberger:2011et}
R.~H. Brandenberger, \emph{{String Gas Cosmology: Progress and Problems}},
  \href{https://doi.org/10.1088/0264-9381/28/20/204005}{\emph{Class. Quant.
  Grav.} {\bfseries 28} (2011) 204005},
  [\href{https://arxiv.org/abs/1105.3247}{{\ttfamily 1105.3247}}].

\bibitem{Brandenberger:2008nx}
R.~H. Brandenberger, \emph{{String Gas Cosmology}},  pp.~193--230, 8, 2008,
  \href{https://arxiv.org/abs/0808.0746}{{\ttfamily 0808.0746}}.

\bibitem{Battefeld:2005av}
T.~Battefeld and S.~Watson, \emph{{String gas cosmology}},
  \href{https://doi.org/10.1103/RevModPhys.78.435}{\emph{Rev. Mod. Phys.}
  {\bfseries 78} (2006) 435--454},
  [\href{https://arxiv.org/abs/hep-th/0510022}{{\ttfamily hep-th/0510022}}].

\bibitem{Patil:2004zp}
S.~P. Patil and R.~Brandenberger, \emph{{Radion stabilization by stringy
  effects in general relativity}},
  \href{https://doi.org/10.1103/PhysRevD.71.103522}{\emph{Phys. Rev. D}
  {\bfseries 71} (2005) 103522},
  [\href{https://arxiv.org/abs/hep-th/0401037}{{\ttfamily hep-th/0401037}}].

\bibitem{Patil:2005fi}
S.~P. Patil and R.~H. Brandenberger, \emph{{The Cosmology of massless string
  modes}}, \href{https://doi.org/10.1088/1475-7516/2006/01/005}{\emph{JCAP}
  {\bfseries 01} (2006) 005},
  [\href{https://arxiv.org/abs/hep-th/0502069}{{\ttfamily hep-th/0502069}}].

\bibitem{Watson:2003gf}
S.~Watson and R.~Brandenberger, \emph{{Stabilization of extra dimensions at
  tree level}},
  \href{https://doi.org/10.1088/1475-7516/2003/11/008}{\emph{JCAP} {\bfseries
  11} (2003) 008}, [\href{https://arxiv.org/abs/hep-th/0307044}{{\ttfamily
  hep-th/0307044}}].

\bibitem{Watson:2004aq}
S.~Watson, \emph{{Moduli stabilization with the string Higgs effect}},
  \href{https://doi.org/10.1103/PhysRevD.70.066005}{\emph{Phys. Rev. D}
  {\bfseries 70} (2004) 066005},
  [\href{https://arxiv.org/abs/hep-th/0404177}{{\ttfamily hep-th/0404177}}].

\bibitem{Brandenberger:2005bd}
R.~Brandenberger, Y.-K.~E. Cheung and S.~Watson, \emph{{Moduli stabilization
  with string gases and fluxes}},
  \href{https://doi.org/10.1088/1126-6708/2006/05/025}{\emph{JHEP} {\bfseries
  05} (2006) 025}, [\href{https://arxiv.org/abs/hep-th/0501032}{{\ttfamily
  hep-th/0501032}}].

\bibitem{Danos:2008pv}
R.~J. Danos, A.~R. Frey and R.~H. Brandenberger, \emph{{Stabilizing moduli with
  thermal matter and nonperturbative effects}},
  \href{https://doi.org/10.1103/PhysRevD.77.126009}{\emph{Phys. Rev. D}
  {\bfseries 77} (2008) 126009},
  [\href{https://arxiv.org/abs/0802.1557}{{\ttfamily 0802.1557}}].

\bibitem{Mishra:2011fc}
S.~Mishra, W.~Xue, R.~Brandenberger and U.~Yajnik, \emph{{Supersymmetry
  Breaking and Dilaton Stabilization in String Gas Cosmology}},
  \href{https://doi.org/10.1088/1475-7516/2012/09/015}{\emph{JCAP} {\bfseries
  09} (2012) 015}, [\href{https://arxiv.org/abs/1103.1389}{{\ttfamily
  1103.1389}}].

\bibitem{Gasperini:1992em}
M.~Gasperini and G.~Veneziano, \emph{{Pre - big bang in string cosmology}},
  \href{https://doi.org/10.1016/0927-6505(93)90017-8}{\emph{Astropart. Phys.}
  {\bfseries 1} (1993) 317--339},
  [\href{https://arxiv.org/abs/hep-th/9211021}{{\ttfamily hep-th/9211021}}].

\bibitem{Tseytlin:1991xk}
A.~A. Tseytlin and C.~Vafa, \emph{{Elements of string cosmology}},
  \href{https://doi.org/10.1016/0550-3213(92)90327-8}{\emph{Nucl. Phys. B}
  {\bfseries 372} (1992) 443--466},
  [\href{https://arxiv.org/abs/hep-th/9109048}{{\ttfamily hep-th/9109048}}].

\bibitem{Gasperini:2002bn}
M.~Gasperini and G.~Veneziano, \emph{{The Pre - big bang scenario in string
  cosmology}}, \href{https://doi.org/10.1016/S0370-1573(02)00389-7}{\emph{Phys.
  Rept.} {\bfseries 373} (2003) 1--212},
  [\href{https://arxiv.org/abs/hep-th/0207130}{{\ttfamily hep-th/0207130}}].

\bibitem{Duff:1989tf}
M.~Duff, \emph{{Duality Rotations in String Theory}},
  \href{https://doi.org/10.1016/0550-3213(90)90520-N}{\emph{Nucl. Phys. B}
  {\bfseries 335} (1990) 610}.

\bibitem{Tseytlin:1990hn}
A.~A. Tseytlin, \emph{{Duality symmetric string theory and the cosmological
  constant problem}},
  \href{https://doi.org/10.1103/PhysRevLett.66.545}{\emph{Phys. Rev. Lett.}
  {\bfseries 66} (1991) 545--548}.

\bibitem{Kugo:1992md}
T.~Kugo and B.~Zwiebach, \emph{{Target space duality as a symmetry of string
  field theory}}, \href{https://doi.org/10.1143/PTP.87.801}{\emph{Prog. Theor.
  Phys.} {\bfseries 87} (1992) 801--860},
  [\href{https://arxiv.org/abs/hep-th/9201040}{{\ttfamily hep-th/9201040}}].

\bibitem{Siegel:1993th}
W.~Siegel, \emph{{Superspace duality in low-energy superstrings}},
  \href{https://doi.org/10.1103/PhysRevD.48.2826}{\emph{Phys. Rev. D}
  {\bfseries 48} (1993) 2826--2837},
  [\href{https://arxiv.org/abs/hep-th/9305073}{{\ttfamily hep-th/9305073}}].

\bibitem{Hull:2009mi}
C.~Hull and B.~Zwiebach, \emph{{Double Field Theory}},
  \href{https://doi.org/10.1088/1126-6708/2009/09/099}{\emph{JHEP} {\bfseries
  09} (2009) 099}, [\href{https://arxiv.org/abs/0904.4664}{{\ttfamily
  0904.4664}}].

\bibitem{Aldazabal:2013sca}
G.~Aldazabal, D.~Marques and C.~Nunez, \emph{{Double Field Theory: A
  Pedagogical Review}},
  \href{https://doi.org/10.1088/0264-9381/30/16/163001}{\emph{Class. Quant.
  Grav.} {\bfseries 30} (2013) 163001},
  [\href{https://arxiv.org/abs/1305.1907}{{\ttfamily 1305.1907}}].

\bibitem{Brandenberger:2017umf}
R.~Brandenberger, R.~Costa, G.~Franzmann and A.~Weltman, \emph{{Point particle
  motion in double field theory and a singularity-free cosmological solution}},
  \href{https://doi.org/10.1103/PhysRevD.97.063530}{\emph{Phys. Rev. D}
  {\bfseries 97} (2018) 063530},
  [\href{https://arxiv.org/abs/1710.02412}{{\ttfamily 1710.02412}}].

\bibitem{Brandenberger:2018xwl}
R.~Brandenberger, R.~Costa, G.~Franzmann and A.~Weltman, \emph{{Dual spacetime
  and nonsingular string cosmology}},
  \href{https://doi.org/10.1103/PhysRevD.98.063521}{\emph{Phys. Rev. D}
  {\bfseries 98} (2018) 063521},
  [\href{https://arxiv.org/abs/1805.06321}{{\ttfamily 1805.06321}}].

\bibitem{Brandenberger:2018bdc}
R.~Brandenberger, R.~Costa, G.~Franzmann and A.~Weltman, \emph{{T-dual
  cosmological solutions in double field theory}},
  \href{https://doi.org/10.1103/PhysRevD.99.023531}{\emph{Phys. Rev. D}
  {\bfseries 99} (2019) 023531},
  [\href{https://arxiv.org/abs/1809.03482}{{\ttfamily 1809.03482}}].

\bibitem{Bernardo:2019pnq}
H.~Bernardo, R.~Brandenberger and G.~Franzmann, \emph{{$T$-dual cosmological
  solutions in double field theory. II.}},
  \href{https://doi.org/10.1103/PhysRevD.99.063521}{\emph{Phys. Rev. D}
  {\bfseries 99} (2019) 063521},
  [\href{https://arxiv.org/abs/1901.01209}{{\ttfamily 1901.01209}}].

\bibitem{Hohm:2019jgu}
O.~Hohm and B.~Zwiebach, \emph{{Duality invariant cosmology to all orders in
  $\alpha$'}}, \href{https://doi.org/10.1103/PhysRevD.100.126011}{\emph{Phys.
  Rev. D} {\bfseries 100} (2019) 126011},
  [\href{https://arxiv.org/abs/1905.06963}{{\ttfamily 1905.06963}}].

\bibitem{Bernardo:2019bkz}
H.~Bernardo, R.~Brandenberger and G.~Franzmann, \emph{{O$(d,d)$ covariant
  string cosmology to all orders in $\alpha^{\prime}$}},
  \href{https://doi.org/10.1007/JHEP02(2020)178}{\emph{JHEP} {\bfseries 02}
  (2020) 178}, [\href{https://arxiv.org/abs/1911.00088}{{\ttfamily
  1911.00088}}].

\bibitem{Bernardo:2020zlc}
H.~Bernardo and G.~Franzmann, \emph{{$\alpha'$-Cosmology: solutions and
  stability analysis}},
  \href{https://doi.org/10.1007/JHEP05(2020)073}{\emph{JHEP} {\bfseries 05}
  (2020) 073}, [\href{https://arxiv.org/abs/2002.09856}{{\ttfamily
  2002.09856}}].

\bibitem{Bernardo:2020nol}
H.~Bernardo, R.~Brandenberger and G.~Franzmann, \emph{{String Cosmology
  backgrounds from Classical String Geometry}},
  \href{https://arxiv.org/abs/2005.08324}{{\ttfamily 2005.08324}}.

\bibitem{Brandenberger:2019jbs}
R.~Brandenberger, \emph{{Fundamental Physics, the Swampland of Effective Field
  Theory and Early Universe Cosmology}},  in \emph{{11th International
  Symposium on Quantum Theory and Symmetries}}, 11, 2019,
  \href{https://arxiv.org/abs/1911.06058}{{\ttfamily 1911.06058}}.

\bibitem{Bedroya:2019snp}
A.~Bedroya and C.~Vafa, \emph{{Trans-Planckian Censorship and the Swampland}},
  \href{https://arxiv.org/abs/1909.11063}{{\ttfamily 1909.11063}}.

\bibitem{Kiefer:1998qe}
C.~Kiefer, D.~Polarski and A.~A. Starobinsky, \emph{{Quantum to classical
  transition for fluctuations in the early universe}},
  \href{https://doi.org/10.1142/S0218271898000292}{\emph{Int. J. Mod. Phys. D}
  {\bfseries 7} (1998) 455--462},
  [\href{https://arxiv.org/abs/gr-qc/9802003}{{\ttfamily gr-qc/9802003}}].

\bibitem{Bedroya:2019tba}
A.~Bedroya, R.~Brandenberger, M.~Loverde and C.~Vafa, \emph{{Trans-Planckian
  Censorship and Inflationary Cosmology}},
  \href{https://doi.org/10.1103/PhysRevD.101.103502}{\emph{Phys. Rev. D}
  {\bfseries 101} (2020) 103502},
  [\href{https://arxiv.org/abs/1909.11106}{{\ttfamily 1909.11106}}].

\bibitem{Mukhanov:1990me}
V.~F. Mukhanov, H.~Feldman and R.~H. Brandenberger, \emph{{Theory of
  cosmological perturbations. Part 1. Classical perturbations. Part 2. Quantum
  theory of perturbations. Part 3. Extensions}},
  \href{https://doi.org/10.1016/0370-1573(92)90044-Z}{\emph{Phys. Rept.}
  {\bfseries 215} (1992) 203--333}.

\bibitem{Sasaki:1986hm}
M.~Sasaki, \emph{{Large Scale Quantum Fluctuations in the Inflationary
  Universe}}, \href{https://doi.org/10.1143/PTP.76.1036}{\emph{Prog. Theor.
  Phys.} {\bfseries 76} (1986) 1036}.

\bibitem{Mukhanov:1988jd}
V.~F. Mukhanov, \emph{{Quantum Theory of Gauge Invariant Cosmological
  Perturbations}}, {\emph{Sov. Phys. JETP} {\bfseries 67} (1988) 1297--1302}.

\end{thebibliography}\endgroup

\end{document}